\documentclass[12pt]{article}
\begin{document}
\renewcommand{\thefootnote}{\fnsymbol{footnote}}
\sloppy

\newcommand \be  {\begin{equation}}
\newcommand \bea {\begin{eqnarray} \nonumber }
\newcommand \ee  {\end{equation}}
\newcommand \eea {\end{eqnarray}}

\begin{center}
\centering{\bf \Large Minimizing volatility increases large risks}
\end{center}
\begin{center}
\centering{D. Sornette $^{1,2}$, J.V. Andersen $^{3}$ and P. Simonetti $^4$\\
{\it $^1$ Institute of Geophysics
and Planetary Physics\\ and Department of Earth and Space Science\\ 
University of California, Los Angeles, California 90095\\
$^2$ Laboratoire de
Physique de la Mati\`ere Condens\'ee\\
CNRS UMR6622 and Universit\'e des
Sciences, B.P. 70, Parc Valrose\\ 06108 Nice Cedex 2, France\\
$^3$ Nordic Institute for Theoretical Physics\\
 Blegdamsvej 17, DK-2100 Copenhagen, Denmark\\
$^4$ Department of Physics and Astronomy, University of Southern California\\
Los Angeles, CA 90089-0484
}
}
\end{center}

\vskip 2cm

{\bf Abstract}\,: We introduce a faithful
representation of the heavy tail multivariate distribution of asset returns,
as parsimonous as the Gaussian framework. Using calculation techniques of
functional integration and Feynman diagrams borrowed from particle physics, we
characterize precisely, through its cumulants of high order, the distribution of 
wealth variations of a portfolio composed of an arbitrary mixture of assets.
The portfolio which minimizes the variance, i.e.
the relatively ``small'' risks, often increases larger risks as measured
by higher normalized cumulants and by the Value-at-risk.

\vskip 1cm

\pagebreak

Finance is all about risks and risk is usually quantified by the volatility. 
As is now well recognized, due to the presence of heavy tails and long-range correlations,
the volatility is only an imperfect measure of risk. In principle, the risk 
associated with a given portfolio is fully embedded in the multivariate
distribution of the returns of these assets. Practically, dealing with this multivariate
distribution is a formidable task for both its specification (important
for scenario simulations), for portfolio
optimization and for the control of risks. 
Until now, simpler one-dimensional measures of risks have been developed,
for instance in terms of the Value-at-risk. However, they suffer from 
their reliance on a stable and accurate determination of the covariance matrix of returns,
which is problematic in the presence of heavy tails and of time-varying volatilities and
correlations. A variety of methods have been
also proposed that are however all limited in their domain of application.

Here, we focus our attention upon the ``fat tail'' problem, having in mind
that a large part of the time-varying volatilities and
correlations may result from their unstable determination precisely due to the presence of
non-gaussian effects. Generalization of our ``fractal'' covariance matrix approach 
described below in the spirit of GARCH models is straightforward and will be described elsewhere.

To address the ``fat tail'' problem, we present three important innovations.
First, we develop a new method that provides 
an approximate but faithful representation
of the full multivariable ``fat tail'' distribution of asset returns. 
Second, we adapt
theoretical tools from theoretical physics to calculate precisely the distribution of
returns of the full portfolio. Third, we compare different portfolio optimization
procedures and show that minimizing the variance is not optimal as it may often
increase large risks. We provide the relevant tools for better optimization suitable
to a given risk aversion.

\section{``Fractal'' Covariance Approximation}

Consider two heterogeneous assets, 
such as the US index SP500 and the Swiss Franc (CHF), both
quoted in US dollars. The empirical joint
bivariate distribution of their dayly annualized
returns 
\be
r_i(t) = 250 ~\ln {s_i(t+1) \over s_i(t)}
\label{hfjjkksll}
\ee
 is plotted in Fig.1 for the time interval from Jan. 1971 
to Oct. 1998. $s_i(t)$ is the price at time $t$ valued in
US dollars, where $i=1$ for the SP500 and $i=2$ for the CHF. 
The contour lines define the probability confidence level\,: 
$95\%$ of the events fall inside the domain limited by the outer line.
Thus, there is a $5\%$ probability to observe events falling outside. The
other confidence levels of $90\%$, $50\%$ and $10\%$ are similarly defined.
Fig.~1 also shows the marginal distributions for the SP500 and the CHF in US\$.
The abcissa axis are the same as for the bivariate representation so that the projection
from the bivariate to the monovariate distributions is highlighted. The ordinate of the
marginal distributions uses a
logarithmic scale\,: a linear plot then qualifies an exponential distribution. One 
can observe that, while the distributions are not far from an exponential, they 
exhibit a slightly upward curvature in the tails indicating a slightly more heavy tail than
the exponential. 

\subsection{Contracting maps as a new quantification of departure from Gaussian}

Let us call $F_{1}(r)$ and $F_{2}(r)$ their cumulative marginal distributions,
giving the probability that the return be less than $r$. Let us introduce the transformation
$r_{1} \to y_{1}$ and $r_{2} \to y_{2}$ which transforms
$F_{1}(r_{1})$ and $F_{2}(r_{2})$ into Gaussian distributions with unit variance. 
By the conservation of probabilities, this reads
\be
F(r_{1,2}) = {1 \over 2} \biggl[1 + {\rm erf}\left({y_{1,2} \over \sqrt{2}} \right)\biggl]~,
\label{rffcd}
\ee
where ${\rm erf}(y)$ is the error function. We can rewrite it in order to make
explicit the nonlinear transformation from the $r$ variables to the $y$ variables\,:
\be
y_{1,2}(r_{1,2})=\sqrt{2}\,{\rm erf}^{-1}\left(2F_{1,2}(r_{1,2})-1 \right)
\label{cujqqllql}
\ee
where erf$^{-1}$ is the inverse of the error function.  The Gaussian $y$ variables,
together with the nonlinear transformation (\ref{cujqqllql}), 
embody fully and {\it with no approximation} the heavy tail nature of the marginal
distributions.

As an analytic illustration, consider stretched exponential (Weibull) distributions
\cite{Lahsor} of the form 
\be
P(r) \equiv {dF \over dr} = {c \over 2\sqrt{\pi}} |r|^{{c \over 2}-1} ~e^{-|{r \over r_0}|^c}~.
\label{dfgjksl}
\ee
This function provides a reasonable fit to the distributions $F_{1}(r_{1})$ and $F_{2}(r_{2})$,
especially in the tails as shown in Fig.~1, with $A_1 = 4500$, $c_1 =0.7$, $r_{01} = 0.79$ and 
$A_2 = 700$, $c_2 =1.1$, $r_{02} = 2.13$.  It is clear that the tails
are much ``fatter'' than for a Gaussian.

In this case, the change of variable (\ref{cujqqllql}) can be written, using a slight change
of normalization, as
\be
y_i(t) = \rm{sign}(r_i(t))~|r_i(t)|^{c_i \over 2}~.
\label{rtgbn,jj}
\ee
The distribution of $y_i$ is Gaussian with a variance now equal to $V_{ii} = (r_{i0})^{c_i}$.

We stress that the transformation (\ref{cujqqllql}) is exact and valid for any 
distribution. It will be used in the simulations. In contrast, 
expression (\ref{rtgbn,jj}) is exact only for Weibull
distributions. It is found to provide a good approximation of the tails of the 
return distributions. It is used below to present the novel theoretical approach.

Fig.~2 shows $y_{1}$ as
a function of $r_{1}$ and $y_{2}$ as a function of $r_{2}$ from the data and the
comparison with (\ref{rtgbn,jj}) using the same parameters as above and shown in Fig.~1.
The negative
returns have been folded back to the positive quadrant. The fits 
with expression (\ref{rtgbn,jj}) shown in Fig.~2 are
good for the large values, while deviations for small returns indicate that 
the departure from a Gaussian is less strong in the center of the distributions. This plot
provides a novel quantification of the departure from a Gaussian.
The downward curvatures result from the fact that 
the tails of the distribution are ``fatter'' than a Gaussian: the $r \to y$
transformation is thus a contracting map. 

\subsection{Optimal multivariate distributions for ``fat tails''} 

To put our next step into its relevant context, we recall that 
multivariate Gaussian distributions have played and still play a key role, not
only because they are convenient to use, but also because they
are optimal in an information theoretical sense\,: with the only prior information
of the covariance matrix, they contain the least possible assumptions, in other words
they are the most likely representation of the data. As already pointed out, they are however
inconsistent with the presence of heavy tails and non-normal dependence. 
In this light, we can now capitalize upon the transformation (\ref{cujqqllql}) and include
the information on the heavy tails to better characterize the multivariate
asset return distributions. In this goal, the dependence between the assets 
is characterized by the covariance matrix $V$ of the
transformed Gaussian variables $y$'s\,:
\be
V = \langle {\bf Y} ~{\bf Y}^T \rangle - \langle {\bf Y} \rangle ~\langle {\bf Y}^T \rangle~,
\label{klgoioc}
\ee
where $\langle {\bf Y} \rangle$ denotes the expectation of ${\bf Y}$ and ${\bf Y}$ is the unicolumn
matrix with elements $y_{1}$ and $y_{2}$. This generalizes straightforwardly
for a larger number $N$ of assets. For stretched exponential variables for which 
the relation (\ref{rtgbn,jj}) holds, the definition (\ref{klgoioc}) leads to the 
covariance elements
\be
V_{ij} =  \langle \biggl(\rm{sign}(r_{i})|r_{i}|^{c \over 2}\biggl)~\biggl(
\rm{sign}(r_{j})~|r_{j}|^{c \over 2}\biggl) \rangle -
 \langle \biggl(\rm{sign}(r_{i})|r_{i}|^{c \over 2}\biggl) \rangle ~\langle \biggl(
\rm{sign}(r_{j})~|r_{j}|^{c \over 2}\biggl) \rangle
\ee
Note that the essential information on the sign of the returns is kept while a
fractional power of their amplitudes is taken, hence the term ``fractal'' covariance matrix
(that we keep even for the general case to refer to the contracting nature of the 
$r \to y$ mapping).
$V_{ij}$ has a faster convergence rate for sparse data and is better
behaved statistically than the usual covariance matrix 
since it is less sensitive to large fluctuations due to the 
small power $c/2$. As a test, we have verified that
the normalized correlation coefficient
$\rho_{V} \equiv {V_{12} \over \sqrt{V_{11} V_{22}}}$
for the covariance matrix for the $y_1$ and $y_2$ variables is significantly more stable than the
usual correlation coefficient $\rho_v \equiv {v_{12} \over \sqrt{v_{11} v_{22}}}$
for the covariance matrix of the returns $r_1$ and $r_2$, as a function of time
in running windows of various sizes.
The introduction of ARCH models and their generalizations has been motivated by
the observed non-stationarity of the usual covariance matrix \cite{ARCH}. 
The improved stability of $\rho_{V}$ suggests that 
this non-stationarity results in part from the inadequacy of the covariance matrix
to provide an efficient characterization of the asset risk profiles, resulting from the
presence of ``fat tails''. Our new approach directly addresses this problem.

Conditionned only on the measurement (\ref{klgoioc}) of the ``fractal'' covariance matrix $V$,
the most likely representation of the time series becomes the usual Gaussian multivariate
distribution in terms of the $y$ variables\,:
\be
  {\hat P}({\bf Y}) = (2\pi)^{-N/2}\,\vert V\vert^{-1/2} 
      \exp\left({-{{\textstyle{1\over2}}}~ ({\bf Y}^T - 
      \langle {\bf Y}^T \rangle) ~~ V^{-1} ~~ ({\bf Y} - \langle {\bf Y} \rangle)} \right)~,
  \label{fjjlqlql}
\ee
where $\vert V \vert$ is the determinant of $V$. We stress that this parameterization is
fundamentally different from the usual Gaussian approximation on the price returns $r$.
To get the implied multivariate distribution $P({\bf R})$ in terms of the return variables
${\bf R}^T = \{r_{1}, r_{2}\}$, 
we use the identity $P({\bf R}) = {\hat P}({\bf Y}) ~{d{\bf Y} \over d{\bf R}}$, where
${d{\bf Y} \over d{\bf R}}$
is the jacobian of the transformation from ${\bf R} \to {\bf Y}$\,:
\be
P({\bf R})= \vert V \vert^{-1/2}
          \exp\left(-{{\textstyle{1\over2}}}\,({\bf Y}^T -\langle {\bf Y}^T \rangle)
          ~~ (V^{-1}- I)~~({\bf Y} - \langle {\bf Y} \rangle)
              \right)
              \prod_{j=1}^N {dF_j \over dr_j}(r_j)~,
\label{qqqlllq}
\ee
where $V$ is again the covariance matrix for ${\bf Y}$ (i.e. the ``fractal''
covariance matrix for ${\bf R}$) and $I$ is identity matrix. Changing the normalization
as in the change of variable (\ref{rtgbn,jj}) leads to the same form (\ref{qqqlllq})
except for the identity matrix $I$ being changed into the diagonal matrix of elements 
$V_{ii} = (r_{0i})^{c_i}$.
This representation is exact for arbitrary uncorrelated variables, in which case $V=I$.
It is also exact for a Gaussian distribution modified by
monotonic one-dimensional variable transformations for any 
number of variables, or equivalently by multiplication by a non-negative separable function.
This  method has recently been independently
introduced in the context of multivariate distributions of particle physics data \cite{Karlen}.

Fig.~3 presents the bivariate distribution ${\hat P}({\bf Y})$ obtained from Fig.~1 using 
the transformation (\ref{cujqqllql}) as well as the corresponding Gaussian marginal
distributions. The contour lines are defined as in Fig.~1.
Note their smooth elliptic shape that contrast with the diamond shape shown in Fig.~1. 
The principal axis of the ellipses are almost
perfectly along the $y_1, y_2$ axis, a signature of the weak ``fractal'' 
correlation between the SP500 and the CHF. In the limit of absence of correlation,
the ratio of the small over large principal axis is  equal to $\sqrt{V_{11} \over V_{22}}$.

As a simple and efficient ``goodness of fit'' test for 
the reliability of this representation (\ref{qqqlllq}), we have studied 
the fraction of events (points) shown in Fig.~3 within an ellipse
of equation $\chi^2 = ({\bf Y}^T - \langle {\bf Y}^T \rangle) \,V^{-1}\,
({\bf Y} - \langle {\bf Y} \rangle)$ as a function of the $\chi^2$ density
$({1/2})~e^{-\chi^2/2}$ for two degrees of freedom. 
We observe a very straight bisector line which qualifies the multivariate Gaussian 
representation (\ref{qqqlllq}).  Varying $\chi^2$ from 
$0$ to $1$ spans the distribution from the
small most probable returns to the large least probable returns.

\section{Characterization of portfolios}

\subsection{Empirical investigation}

We can now capitalize upon the rather good stationarity properties of the representation
of the bivariate distributions provided by (\ref{qqqlllq}) and use
this information to optimize portfolios and characterize risks. 
Consider a portfolio investing a fixed fraction $p$ of 
its wealth $W$ in the SP500 and the remaining fraction $1-p$ in the CHF. 
Using the historical time series, 
we construct numerically the time series $W(t)$ from the recursion 
\be
W(t+1) = p W(t) s_1(t) + (1-p) W(t) s_2(t)
\label{hgjjjjjd}
\ee
which ensures that $p$ is fixed. The annualized dayly return $r_W$ of $W(t)$ is
defined by $r_W(t) = 250~ \ln {W(t+1) \over W(t)}$.
Fig.~4 shows the dependence as a function of $p$ of the variance 
\be
C_2 \equiv \langle (r_W - \langle r_W \rangle)^2 \rangle
\ee
and of the kurtosis 
\be
\kappa \equiv {C_4 \over C_2^2} = {\langle (r_W - \langle r_W \rangle)^4 \rangle 
\over \langle (r_W - \langle r_W \rangle)^2 \rangle^2} - 3~,
\ee
of the dayly portfolio returns. The kurtosis quantifies the deviation from a Gaussian
distribution and provides a measure for the degree of ``fatness'' of the tails, i.e.
a measure of the ``large'' risks. Taking into account only the variance and the kurtosis 
and neglecting all higher order cumulants, a distribution can be approximated by the
following expression valid for small kurtosis \cite{PhysicaA}
\be
P_(r_W) \simeq \exp \biggl[-{(r_W-\langle r_W \rangle)^2 \over 2 C_2}
\biggl(1 - {5 \kappa \over 12} {(r_W-\langle r_W \rangle)^2 \over C_2}\biggl)\biggl] ~ . 
\label{gaussddd}
\ee
The negative sign
of the correction proportional to $\kappa$ means that large deviations are more probable
than extrapolated from the Gaussian approximation.
For a typical fluctuation $|S - \langle S| \rangle \sim \sqrt{C_2}$, the relative size of the 
correction in the exponential is ${5 \kappa \over 12}$. For the large values of
$\kappa$ found below this approximation (\ref{gaussddd}) break down and the deviation 
from a Gaussian is much more dramatic.

As seen in Fig.~4, the variance has a well-defined quadratic minimum at $p_V = 0.375$. 
The kurtosis has a S-shape with two local minima at $p_{\kappa2} = -0.405$ (absolute
minimum) and $p_{\kappa1} = 0.125$ (local minimum). The table gives the corresponding
variance $C_2$ and kurtosis $\kappa$ for these three portfolios and for the benchmark
$p_B = 0.5$. 

\vskip 0.5cm
\begin{table*}[h]
\begin{center}
\begin{tabular}{|c|c|c|c|c|c|c|c|c|} \hline
$p$ & $C_2$ & $\kappa$ & $r_{\ell}$ & $c$ & VaR ($20$ days) & VaR ($10$ years)\\ \hline
$p_B = 0.5$ & $2.42$ & $19.9$ & $1.0$ & $0.75$ & $-3.77$ & $-19.4$\\ \hline
$p_V = 0.375$ & $2.28$ & $9.53$ & $1.77$ & $1.09$ & $-4.41$ & $-13.6$ \\ \hline
$p_{\kappa1} = 0.125$ & $2.85$ & $4.20$ & $3.44$ & $1.73$ & $-6.12$ & $-12.4$\\ \hline
$p_{\kappa2} = -0.405$ & $7.77$ & $3.92$ & $4.39$ & $1.35$ & $-9.19$ & $-22.8$\\ \hline
\end{tabular}
\end{center}
Table: $p$ (resp. $1-p$) is the weight in value invested in the SP500 (resp. CHF). $C_2$
(resp. $\kappa$) is the variance (resp. kurtosis) of the distribution of returns of the
portfolios. $r_{\ell}$ and $c$ are the scale and exponent of the Weibull fit to their tail.
The last two columns report the calculated Value-at-Risk at the $95\%$ and $99.96\%$ 
confidence levels.
\end{table*}
\vskip 0.5cm

The conclusion of this analysis is striking\,: the portfolio with $p_{\kappa1} = 0.125$ has
a variance only $25\%$ higher than that of the minimum variance portfolio
while its kurtosis is smaller than half that of the minimum variance portfolio. It is thus
possible to construct a portfolio which
has about the same degree of ``small'' risks (as measured by the variance) while having
significantly smaller ``large'' risks than would give the standard ``mean-variance''
portfolio approach \cite{Markovitz}.

This result can also be interpreted in a way that highlights the danger of standard
practice\,: minimizing ``small'' risks as quantified by the variance may increase 
(here more than double) the ``large'' risks. In trouble times of large volatity fluctuations,
it is particularly important to recognize this fact. Fig.~5 further exemplifies this
phenomenon by plotting the cumulative distributions $F(r_W)$ for the four 
portfolios in an inverse axis representation, 
corresponding to the so-called Zipf or rank-ordering plot\,: this representation
of the $n$th largest value as a function of its rank $n$ emphasizes the information
in the tail of the distribution. We can collapse the tails of the 
distributions of the four portfolios by choosing suitable pairs of parameters
$c$ and $r_{\ell}$ for each portfolio distribution and by plotting $(r_W/r_{\ell})^c$ 
as a function of $\ln n$\,: this
collapse is the signature that all the tails are approximately of the same 
functional form (\ref{dfgjksl}) and that we have correctly identified the values of the parameters.
The table lists the values of $c$ and $r_{\ell}$ 
that best fit the tail of each portfolio return distribution. The portfolio with 
$p=0.125$ provides the best compromise with a low variance and a low kurtosis:
not surprisingly, the exponent $c$ of its tail is the largest corresponding to the 
faster asymptotic decay (thinnest tail).

\subsection{Theoretical formulation}

We now present briefly how these stylized facts can be rationalized by
a systematic theory based on the representation (\ref{qqqlllq}).
Up to a very good approximation, it is harmless and much simpler to replace
the returns $r_i(t)$ defined in (\ref{hfjjkksll}) by $(s_{i}(t+1) - s_i(t))/s_i(t)$ and, 
over reasonable large time intervals (e.g a year), 
neglect the variation the denominator in comparison
to the variation of the numerator $\delta s_i(t) \equiv s_{i}(t+1) - s_i(t)$.
The dayly wealth variation at time $t$ of a portfolio of $N$ assets reads
\be 
\delta W(t) = \sum_{i=1}^N p_i ~\delta s_i(t) ~,
\label{sumport}
\ee
where $p_i$ is again the weight in value of the $i$th asset
in the portfolio. We normalize the weights $\sum_{i=1}^N p_i = 1$. 
Our strategy is to express the $\delta s_i(t)$ variables as a function of the $y_i(t)$ using
(\ref{cujqqllql}) and calculate directly the distribution $P(\delta W)$ of the dayly
portfolio wealth variations. We stress that $P(\delta W)$ embodies completely all possible
information on risks and in particular embodies the usual volatility and VAR measures.
We illustrate the procedure for the case of 
Weibull distributions for which (\ref{cujqqllql}) reduces to (\ref{rtgbn,jj})\,:
\be 
\delta W(t) = \sum_{i=1}^N p_i ~\rm{sign}(y_i)~ |y_i|^{2 \over c_i}~.
\label{sumpqgort}
\ee
The formal expression for $P(\delta W)$ is
\be
P(\delta W) = C \prod_{i=1}^N \biggl( \int dy_i \biggl) ~ e^{-{1 \over 2} ~{\bf Y}^T V^{-1} {\bf Y} }
~\delta\biggl(\delta W(t) - \sum_{i=1}^N p_i \rm{sign}(y_i)~ |y_i|^{2 \over c_i} \biggl) ~ .
\label{soqglut}
\ee
In order to simplify the notation, we assume that the average price variations are zero. It is easy to
reintroduce non-zero average returns in the formalism.
Taking the Fourier transform of (\ref{soqglut}), we get
\be
{\hat P}(k) = {1 \over (2\pi)^{N/2} \det{V}^{1/2} }
 \prod_{i=1}^N \biggl( \int du_i \biggl)  e^{-{1 \over 2} ~^T {\bf Y} V^{-1} {\bf Y} +
ik~\sum_{i=1}^N ~p_i ~y_i^{q_i} }~ ,
\label{soqqgglurft}
\ee
where $c_i=2/q_i$. We only show the expression (\ref{soqqgglurft}) for the case
where $q_i$ are integers and odd such that the ``interaction'' terms
sign$(y_i)~|y_i|^{2 \over c_i}$ simplify into $y_i^{q_i}$.
Note that the case $q=3$  corresponding to an exponent $c=2/3$ is realistic 
empirically for the SP500 data. Our results below holds for general $q$'s.
Expression (\ref{soqqgglurft}) bears strong ressemblance with quantities that appear
in field theories of particle physics and we have used the relevant ``technology''
to evaluate it.

For $q=1$, i.e. $c=2$, the change of variable (\ref{rtgbn,jj}) is linear, all integrals are gaussian
which yields the standard result that the distribution $P(\delta W)$ is Gaussian with a variance
\be
C_2 = p^T V p~.
\ee
This retrieves the results covered by the standard Markovitz's theory \cite{Markovitz}
at the basis of the CAPM \cite{Merton}. 

Consider now the more general ``heavy tail'' case of arbitrary $q>1$, i.e. $c=2/q<2$. 
For uncorrelated assets, $V$ is diagonal and the multiple integral becomes the product of 
one-dimensional integrals. We have shown \cite{Tobepub} that cumulants of $P(r_W)$ of all
orders can be calculated exactly\,:
\be
C_{2 n}(q) =  \sum_i C(n,q_i)~(p_i^2 v_{ii})^{n}~,
\label{gghJD/D}
\ee
where $C(n,q)$ is a function of $n$ and $q$ \cite{Tobepub}.
We have $C(1,q) = (2^q/\sqrt{\pi}) \Gamma(q+1/2)$ and 
$C(2,q) = (2^{2q}/\sqrt{\pi})~\Gamma(2q+1/2) - (3~2^{2q}/\pi) [\Gamma(q+1/2)]^2$, where
$\Gamma$ is the Gamma function. In this diagonal case, the $q_i$th power
of the variance of $y_i$ is equal to the variance $v_{ii}$ of the $i$th asset
dayly price variation $\delta s_i$, leading to $(V_{ii})^q = v_{ii}$. 
We stress that this expression (\ref{gghJD/D}) is valid even when $q$ is real 
and the interaction term is $\propto {\rm sign}(y_i) |y_i|^q$ and thus
applies to arbitrary Weibull exponential distributions. Odd cumulants are
vanishing due to our restriction to distribution with zero mean. 

It is well-known
that, conditionned on mild regularity conditions, the knowledge of all cumulants 
uniquely determines the distribution function $P(\delta W)$. We have thus been
able to characterize fully in this case all aspects of risks associated to a given
portfolio. Recall that the cumulant $C_2$ is the variance of the portfolio wealth
variation distribution. 
The normalized fourth cumulant $\kappa \equiv {C_4 \over C_2^2}$ is its kurtosis. As already
mentionned, it is zero for
a Gaussian distribution and provides a standard measure of departure from Gaussian.
Higher order cumulants quantify the deviation from a Gaussian further in the tail of the
distribution. 

We have also been able to calculate the cumulants for the correlated case. 
The calculation is significantly more involved and uses
a systematic Feynman diagrammatic procedure \cite{PhysicaA,Tobepub} that has been 
invented in quantum electrodynamics \cite{Diagrammatica}. The results and 
corresponding empirical
tests will be given in \cite{Tobepub}.

This completes our brief summary of our complete {\it analytical} 
determination of the distribution of the portfolio wealth variation for multivariate
correlated fat tail multivariate distributions. Our technique can be
extended to more general asset distributions of the form $P(r) = e^{-f(r)}$, as
long as $f(r) \to +\infty$ for $|r| \to +\infty$ no slower than a power law with positive
exponent. This condition covers all cases of practical interest.

We now use these analytical results to generalize our empirical finding that
minimizing ``small'' risks as quantified by the variance often increases significantly
the ``large'' risks.

\section{Risk quantification}

\subsection{Optimal portfolios}

To keep the presentation simple, we consider the uncorrelated diagonal case (\ref{gghJD/D}).
Being presented the full spectrum of cumulants that quantify all possible measures
of risks, we now determined two ``optimal'' portfolios. 
\begin{itemize}
\item The first porfolio $P_V$ has the smaller variance. The corresponding asset weights 
are found to be\,:
\be
p_1 v_{11} = p_2 v_{22} = \ldots = p_N v_{NN} = {1 \over \sum_i {1 \over v_{ii}}}~,
\label{minvardiag}
\ee
where $v_{ii}$ is the variance of the $i$th asset. The assets contribute to this
portfolio in value inversely proportional to their variance. 

\item The second portfolio $P_K$ has simultaneously the smallest kurtosis ${C_4 \over C_2^2}$
and smallest higher normalized cumulants $\lambda_{2 m} \equiv {C_{2m} \over (C_2)^m}$ for 
$m>2$. The corresponding asset weights are\,:
\be
p_1 v_{11}^{1/2} = p_2 v_{22}^{1/2} = \ldots = p_N v_{NN}^{1/2} = 
{1 \over \sum_i {1 \over v_{ii}^{1/2}}}~.
\label{mincumdiag}
\ee
Since the normalized cumulants  $\lambda_{2 m}$ with $m \geq 2$ measure the deviation
from a Gaussian in the tail, $P_K$ minimizes the large risks.
\end{itemize}

\subsection{Small versus large risk optimization}

The asset weights given by (\ref{mincumdiag}) do not minimize the portfolio variance but do
correspond to the smallest possible large risks. Reciprocally, the asset weights 
given by (\ref{minvardiag}), that minimize the portfolio variance, increase the large risks.
We state two results among several others that we have obtained that generalize
this observation. Let us denote
$$
X_i \equiv {{1 \over v_{ii}^{1/2}} \over 
\sum_{j=1}^N {1 \over v_{jj}^{1/2}}}
$$
the relative inverse risk brought by asset $i$. Let us also 
call $\lambda_{2m}^{(K)}$ (resp. $\lambda_{2m}^{(V)}$)
the normalized cumulant of order $2m$ of the portfolio $P_K$ (resp. $P_V$). Then,
\be
{\lambda_{2m}^{(K)} \over \lambda_{2m}^{(V)}} = {1 \over N^{m-1}} { 
\biggl(\sum_i X_i^2 \biggl)^{m} \over \biggl(\sum_j X_j^{2m}\biggl)}~.             		
\label{mlfgggglm}
\ee
We thus find that $\lambda_{2m}^{(K)}$ is always smaller or equal to 
$\lambda_{2m}^{(V)}$ for $m \geq 2$
for all possible values of $X_i$'s. The equality occurs only for all $X_i$'s being equal to $1/N$,
i.e. for assets with identical variances. This demonstrates that
the weights that minimize the variance {\it increase} the higher normalized cumulants.
It also interesting to compare the portfolio $P_K$ with the benchmark portfolio
$P_{1/N}$ defined by $p_1 = p_2 = ... = p_N = 1/N$. We find
\be
{\rm ratio} \equiv {\lambda_{4}^{(V)} \over \lambda_{4}^{(1/N)}} = { 
\biggl(\sum_i {1 \over v_{ii}^2} \biggl) ~\biggl(\sum_j v_{jj}\biggl)^2 \over
\biggl(\sum_k {1 \over v_{kk}} \biggl)^2 ~ \biggl(\sum_l v_{ll}^2 \biggl)} ~,
\label{mlfgggglqgqdsm}
\ee
Notice that changing all variances $v_{ii}$ into their inverse change the ratio of kurtosis into
its inverse. This implies that, if we find a set of $v_{ii}$'s for which 
the ratio of kurtosis is smaller than one, then the set of the inverses 
$1/v_{ii}$'s gives a ratio of kurtosis larger than one. 
This proves that there are many situations for which minimizing the 
variance of the portfolio may
either increase its kurtosis and therefore its large risks as compared to that of
the benchmark. 

\subsection{Empirical test}

\subsubsection{Kurtosis}

Fig.~6 compares the dependence of the empirical kurtosis shown in Fig.~4 to the 
prediction obtained from Eq.~(\ref{gghJD/D}) of the theory. We use the result
for uncorrelated assets as the coefficient of correlations are small $\rho_v
\approx \rho_V \approx 0.03$. We have checked that taking into account 
the non-zero value of $\rho$ does not change significantly the results.

We show six theoretical curves for all the combinations of the values $c_1 = 0.7$, 
$c_1 = 0.8$ and $c_2 = 1.05$, $c_2 = 1.1$ and $c_2 = 1.15$. 
For relatively large positive (`long') and negative (`short')
 weights $p$ of the SP500, the kurtosis
$\kappa$ is mostly sensitive to the estimation of the exponent $c_1$ of the SP500 return
distribution, because the SP500 has the fatest tail (smallest exponent $c$).
For small values of $p$, the reverse is true and
the portfolio kurtosis is mostly sensitive to the 
exponent $c_2$ of the CHF return distribution. The empirical determination shown in Fig.~4
is replotted as circles. In the domain of $p$ with reasonable variance and kurtosis, we
find a quite good agreement for $c_1 = 0.75, c_2 = 1.15$. The other theoretical curves
provide the range of uncertainty in the kurtosis estimation coming from measurement
errors in the exponents $c$. The main point here is that the theory adequately identifies the
set of portfolios which have small kurtosis and thus small `large risks' and still reasonable
variance (`small risk'). We stress the importance of such precise analytical quantification to
increase the robustess of risk estimators\,: historical data becomes notoriously 
unreliable for medium and large risks for lack of suitable statistics.

\subsubsection{Value-at-Risk}

As a final test, we show how the different portfolios perform with respect to the 
Value-at-Risk (VaR) at different confidence levels.
Recall that the VaR determines the probability of a portfolio of assets losing
a certain amount in a given time period due to adverse market conditions with a 
particular level of confidence $C_L$ \cite{Jorion}. 
For instance, a VaR-measure of one millon dollars at
the $C_L=95\%$ level of confidence implies that total portfolio losses would not exceed
one million dollars more than $1-C_L=5\%$ of the time (i.e. typically one day 
in twenty) over a given holding period. In essence,
VaR provides a measure of extreme events that occur in the lower tail of the portfolio's
return distribution.

We have estimated the VaR for each of the four portfolios both from historical data
and from the stretched exponential model.
For each weight, we constructed the distribution of 
returns $P(r_W)$ obtained from (\ref{hgjjjjjd}) and estimated directly the VaR such that
the fraction of negative returns smaller than VaR is $1-C_L$. Mathematically, this
corresponds to determine the return $r_W$ such that
$F(r_W) =1-C_L$. The corresponding VaRs at the $C_L=95\%$ level are given in the table.
This confidence level corresponds to a typical maximum dayly loss encountered once
every $20$ days.

An independent estimation was performed by using the fits of the
distributions of $r_W$ by stretched exponentials, with the values of $c$ and $r_{\ell}$
reported in the table. That the portfolio distributions can still be considered 
of this form in their tail is validated by an ``extreme deviation'' theorem \cite{Frisch}.
Then, the VaR is solution of
\be
 C_L = {1 \over 2 \alpha} \biggl[1 + {\rm erf}\left({(VaR/r_{\ell})^{c \over 2} \over 
\sqrt{2}} \right)\biggl]~,
\label{rffcqqqssd}
\ee
which has to be solved with respect to $Var$. The additional multiplicative factor
$\alpha \approx 10$ accounts for the empirical fact that the stretched exponential is valid
only in the tail of the distribution. $\alpha$ has been calibrated for one 
confidence level and checked to remain approximately the same for the others. 

This calibration allows us to predict the VaR at higher confidence levels, i.e. for dayly
losses that can typically occur over longer period of times than $20$ days.
For relatively low 
confidence interval like $C_L=95\%$, we find that the VaR for the Variance portfolio $P_V$
$p=0.375$ is significantly smaller than that for the kurtosis portfolio $P_K$ with $p=0.125$.
But since the exponent $c$ of $P_K$ is larger than that of $P_V$, the tail
of $P_K$ is bounded to become thinner and the VaR of the kurtosis portfolio $P_K$ is
bound to become smaller than that of the variance portfolio $P_V$ at high confidence levels.
We calculate that the cross-over occurs approximately at a confidence level of 
$99.93\%$ corresponding to a typical largest dayly loss of about $12\%$ 
occurring once every five years. For larger time horizon, the kurtosis portfolio becomes
better, having a smaller VaR. We show the VaR at the confidence level of $99.96\%$ 
corresponding to the decadal dayly shock, i.e. to the typical largest loss seen 
once every ten years. As expected, the kurtosis portfolio has the smallest VaR.

\pagebreak

\pagebreak

FIGURE CAPTIONS
\vskip 1cm

Fig.~1: Bivariate distribution of the dayly annualized returns of the SP500
US index $1$ and of the CHF $2$ (in US \$) for the time interval from Jan. 1971 
to Oct. 1998. Half of the data is represented for clarity of the figure.
The contour lines define the probability confidence level of 
$95\%$ (outer line), $90\%$, $50\%$ and $10\%$. The upper and left diagrams show
the projected marginal distributions for the SP500 and the CHF in US\$ and their 
fit to (\ref{dfgjksl}). The parameters of the fit are
$A_1 = 4500$, $c_1 =0.7$, $r_{01} = 0.79$ and 
$A_2 = 700$, $c_2 =1.1$, $r_{02} = 2.13$.  

\vskip 0.5cm
Fig.~2: Dependence of the gaussian variables defined by (\ref{cujqqllql})
as a function of the return for the SP500 and CHF data shown in Fig.~1.
The negative returns have been folded back onto the positive quadrant.
The continuous lines are given by (\ref{rtgbn,jj}) with 
$c_1 =0.7$ and $c_2 = 1.1$ respectively for the SP500 $1$ and CHF $2$. 

\vskip 0.5cm
Fig.~3: Bivariate distribution ${\hat P}({\bf Y})$ obtained from Fig.~1 using 
the transformation (\ref{cujqqllql}). The contour lines are defined as in Fig.~1.
The upper and left diagrams show
the corresponding projected marginal distributions, which are gaussian
by construction of the change of variable (\ref{cujqqllql}). Both are fitted by the
continuous line of equation $P_{1,2} =150~\exp(-|y_{1,2}|^2/2)$.

\vskip 0.5cm
Fig.~4: Empirical dependence as a function of $p$ of the variance 
$C_2$ and of the kurtosis $\kappa$ of the distribution of returns
$r_W(t) = 250~ \ln {W(t+1) \over W(t)}$
of a portfolio with a fraction $p$ (resp. 1-p) in value invested in the SP500 index 
(resp. in the CHF), whose total value is given by (\ref{hgjjjjjd}).
The variance has a well-defined quadratic minimum at $p_V = 0.375$. 
The kurtosis has a S-shape with two local minima at $p_{\kappa2} = -0.405$ (absolute
minimum) and $p_{\kappa1} = 0.125$ (local minimum). The table gives the corresponding
variance $c_2$ and kurtosis $\kappa$ for these three portfolios and for the benchmark
$p_B = 0.5$. 

\vskip 0.5cm
Fig.~5: Rescaling of the distributions $P(r_W)$ of returns $r_W$ obtained from the four 
portfolios studied in the table. The rescaling uses for the ordinate the 
reduced variable $(r_W / r_{\ell})^c$ where the exponent $c$ and the characteristic 
return scale $r_{\ell}$ have been determined by a direct fit to each portfolio return
distributions. The abcissa is the rank $n$ of the $n$th largest value plotted
along the ordinate. This rank-ordering plot, which is the same as a cumulative plot, but 
with reversed axis, emphasizes the information contained in the tail.
The symbols correspond to: $+: p=-0.405;  o: p=0.125; *:p=0.375;  x: p=0.5$. The
straight line has equation $7.50-1.16 \ln n$.

\vskip 0.5cm
Fig.~6: Comparison of the empirical kurtosis (circles) shown in Fig.~4 with the
prediction obtained from Eq.~(\ref{gghJD/D}) of the theory. 
The six theoretical curves correspond to all combinations of 
pairs of values $c_1 = 0.7$, $c_1 = 0.9$ and 
$c_2 = 1.15$ (solid  line); $c_2 = 1.1$ (dotted-dashed line) and 
$c_2 = 1.05$ (dotted line).


\begin{thebibliography}{10}

\bibitem{Lahsor} J. Laherr\`ere and D. Sornette, European Physical Journal B 2, 525-539 (1998).

\bibitem{ARCH} R.F. Engle, Econometrica 50, 987 (1982);
 T. Bollerslev, R.Y. Chous and K.F. Kroner, J. Econometrics 52, 5 (1992).
 
\bibitem{Karlen} D. Karlen, Computer in Physics 12, 380-384 (1998).

\bibitem{Markovitz} H. Markovitz, Portfolio selection : Efficient diversification of
investments (John Wiley and Sons, New York, 1959).

\bibitem{Merton} R. C. Merton, Continuous-time finance, (Blackwell, Cambridge,1990).

\bibitem{Tobepub} D. Sornette, J.V. Andersen and P. Simonetti, preprint 1998.

\bibitem{PhysicaA} D. Sornette, Physica A 256, 251-283 (1998).

\bibitem{Diagrammatica} M. Veltman, Diagrammatica, The path to Feynman diagrams
(Cambridge Lecture Notes in Physics, Cambridge UK, 1995).

\bibitem{Jorion} P. Jorion, Value-at-Risk: The New Benchmark for Controlling Derivatives Risk (Irwin
Publishing, Chicago, IL, 1997).

\bibitem{Frisch} U. Frisch and D. Sornette, J. Phys. I France 7, 1155-1171 (1997).


\end{thebibliography}
 
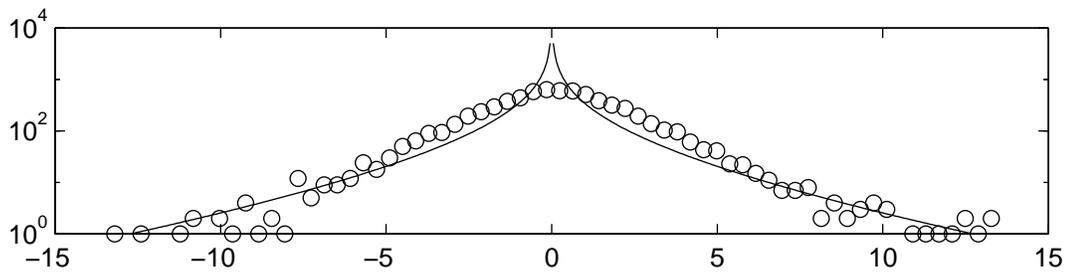
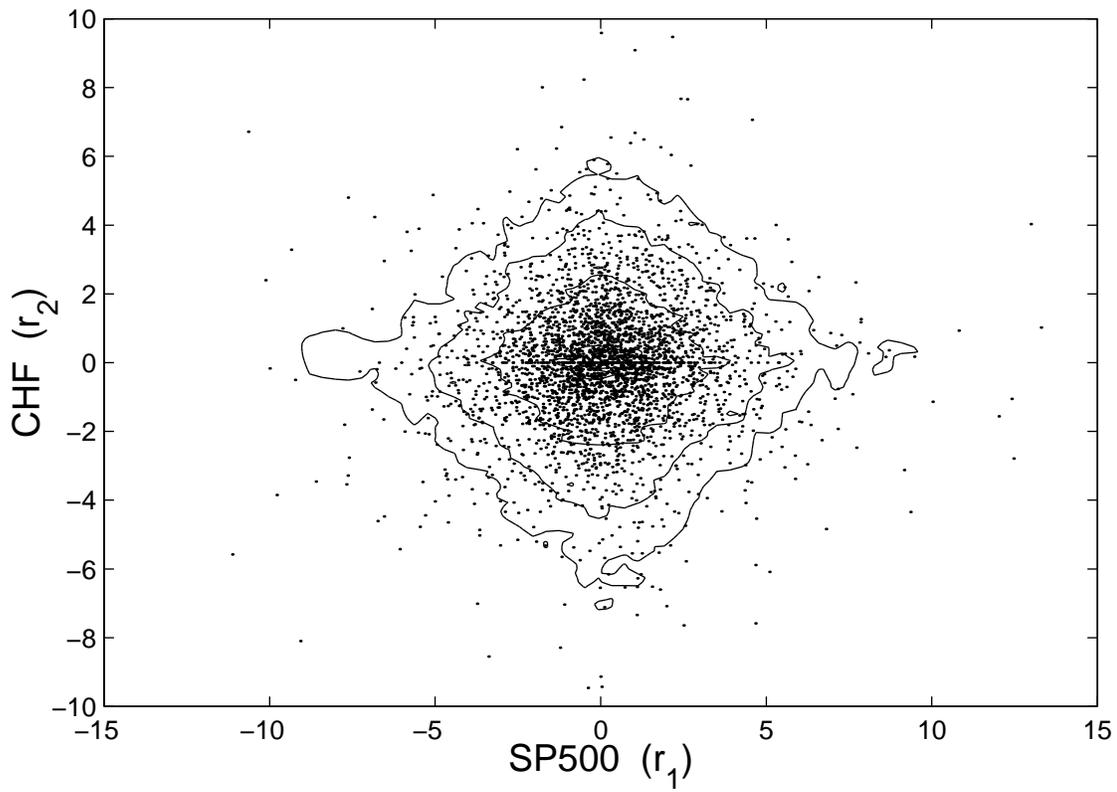
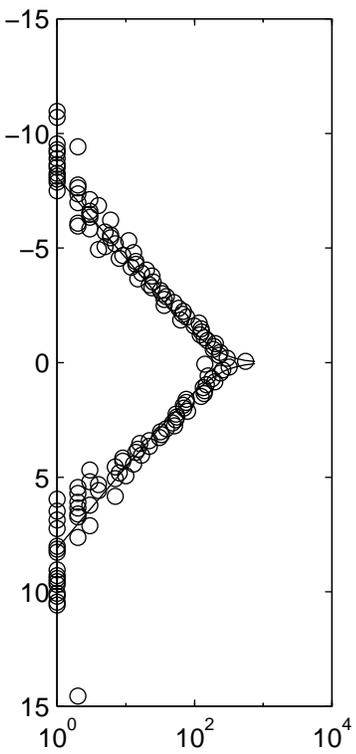
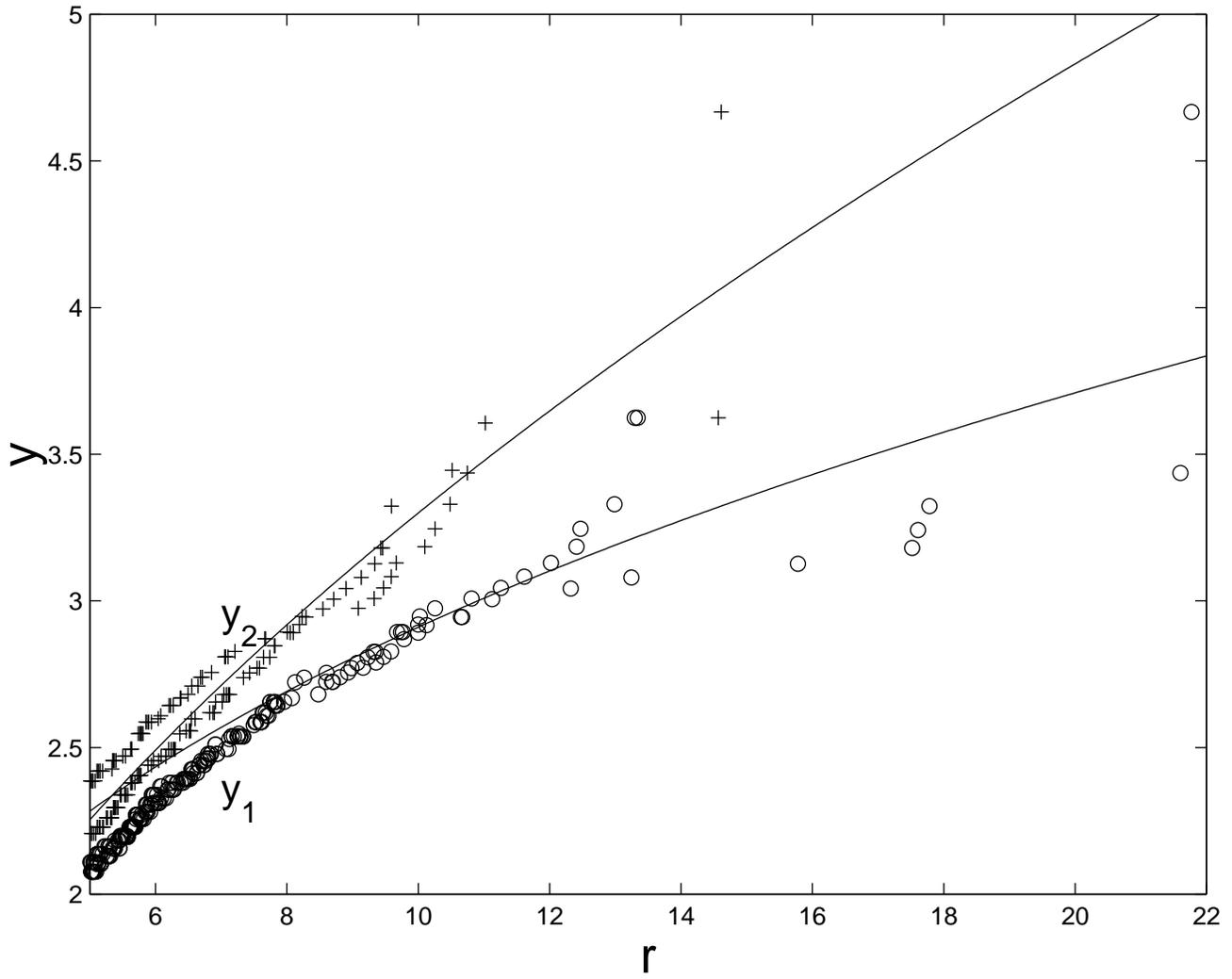
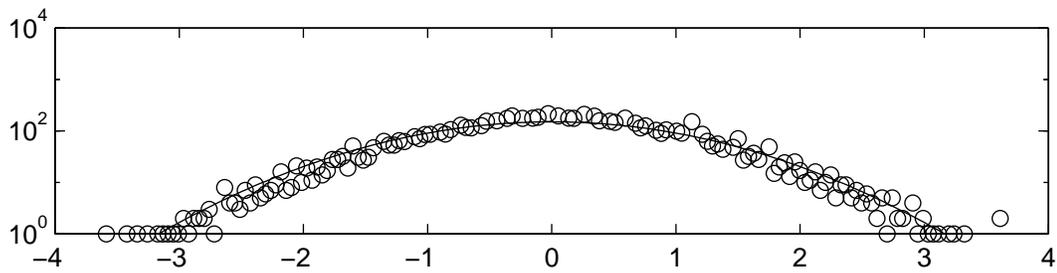
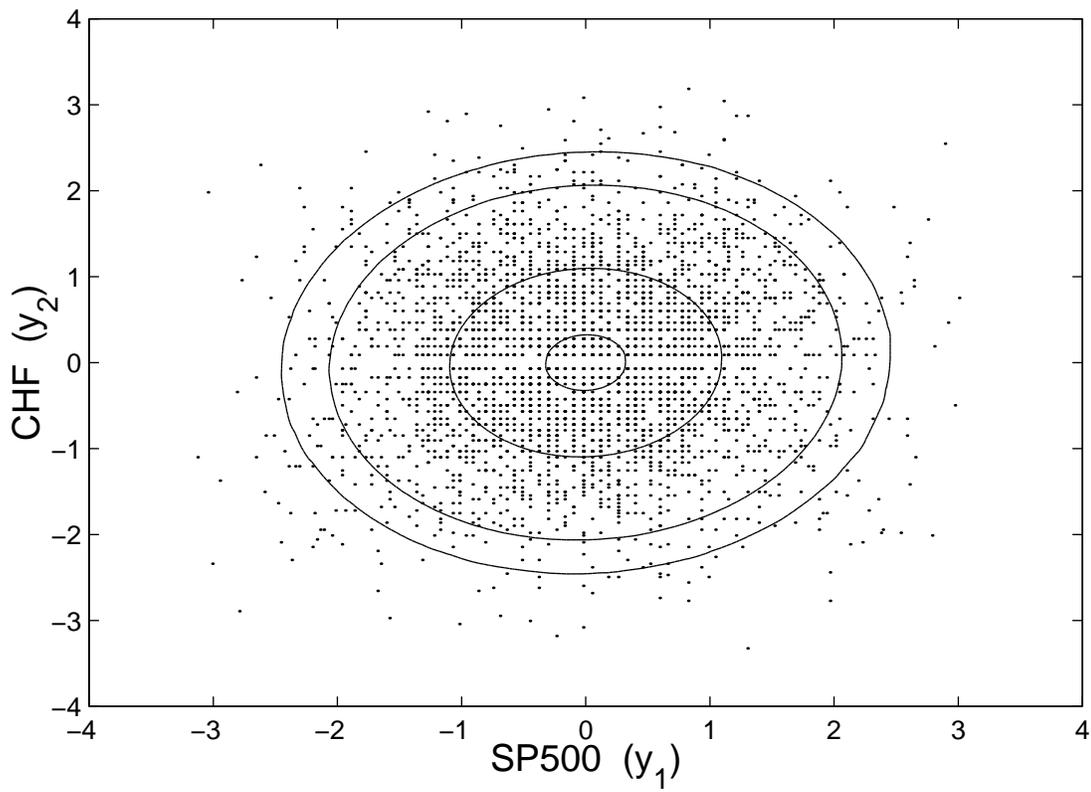
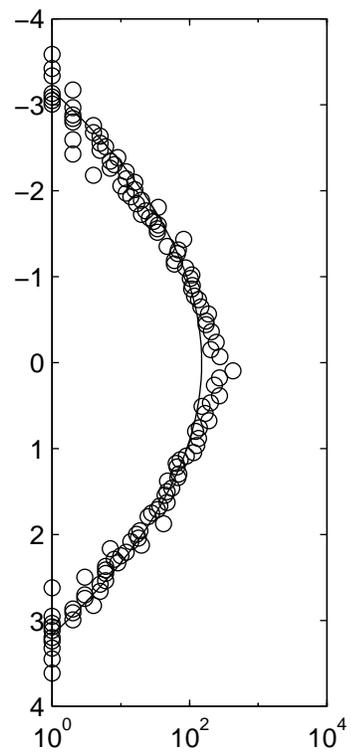
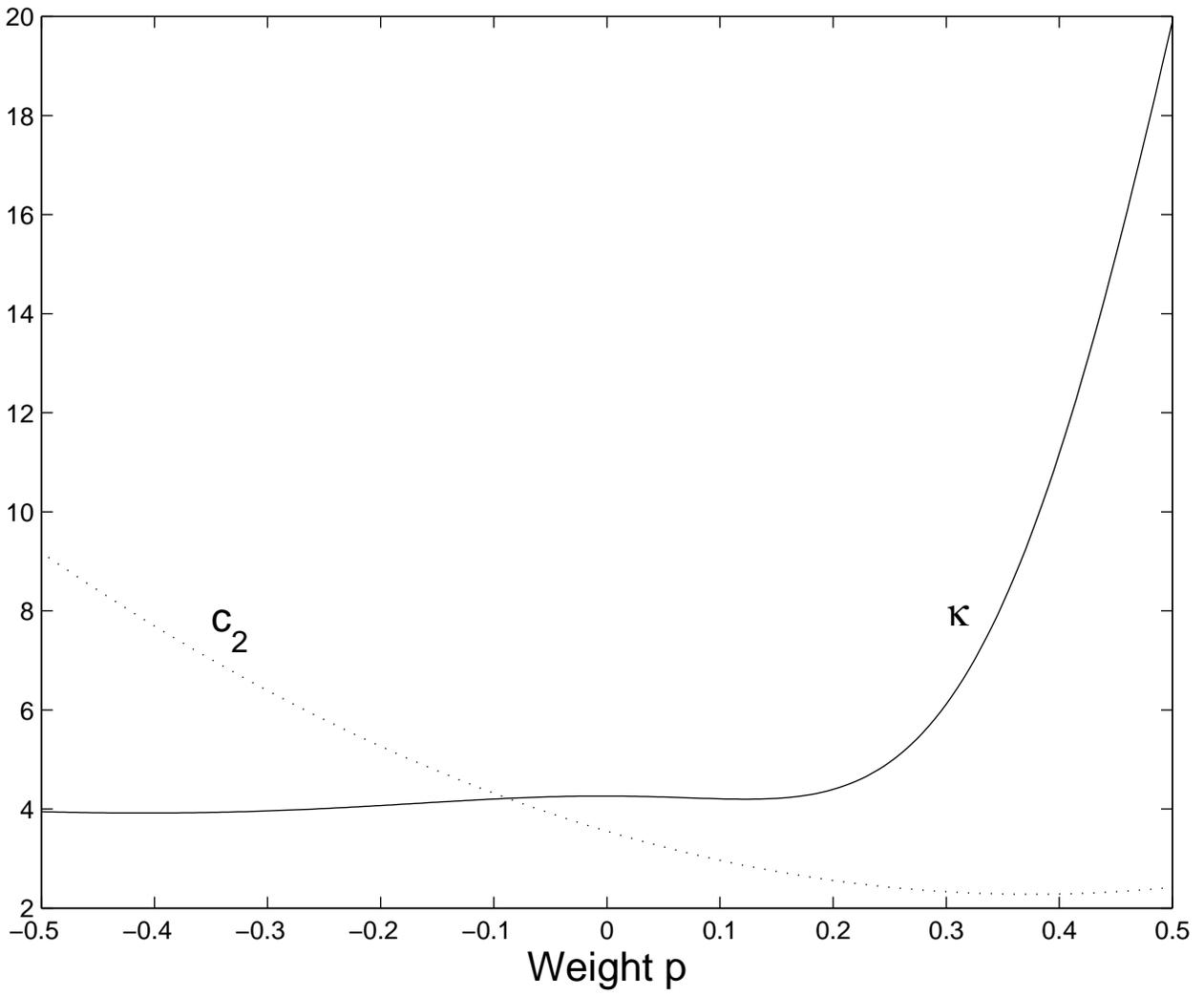
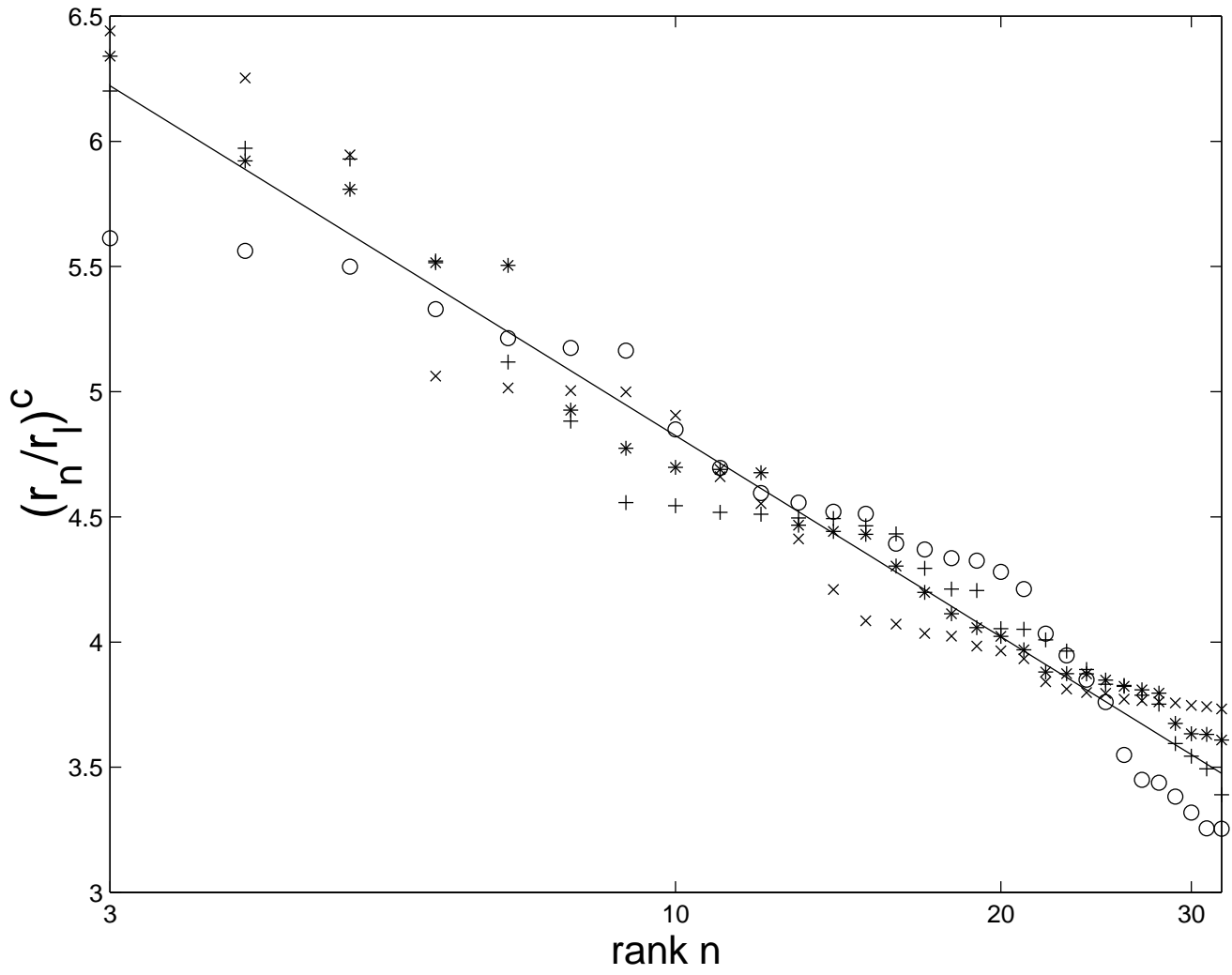
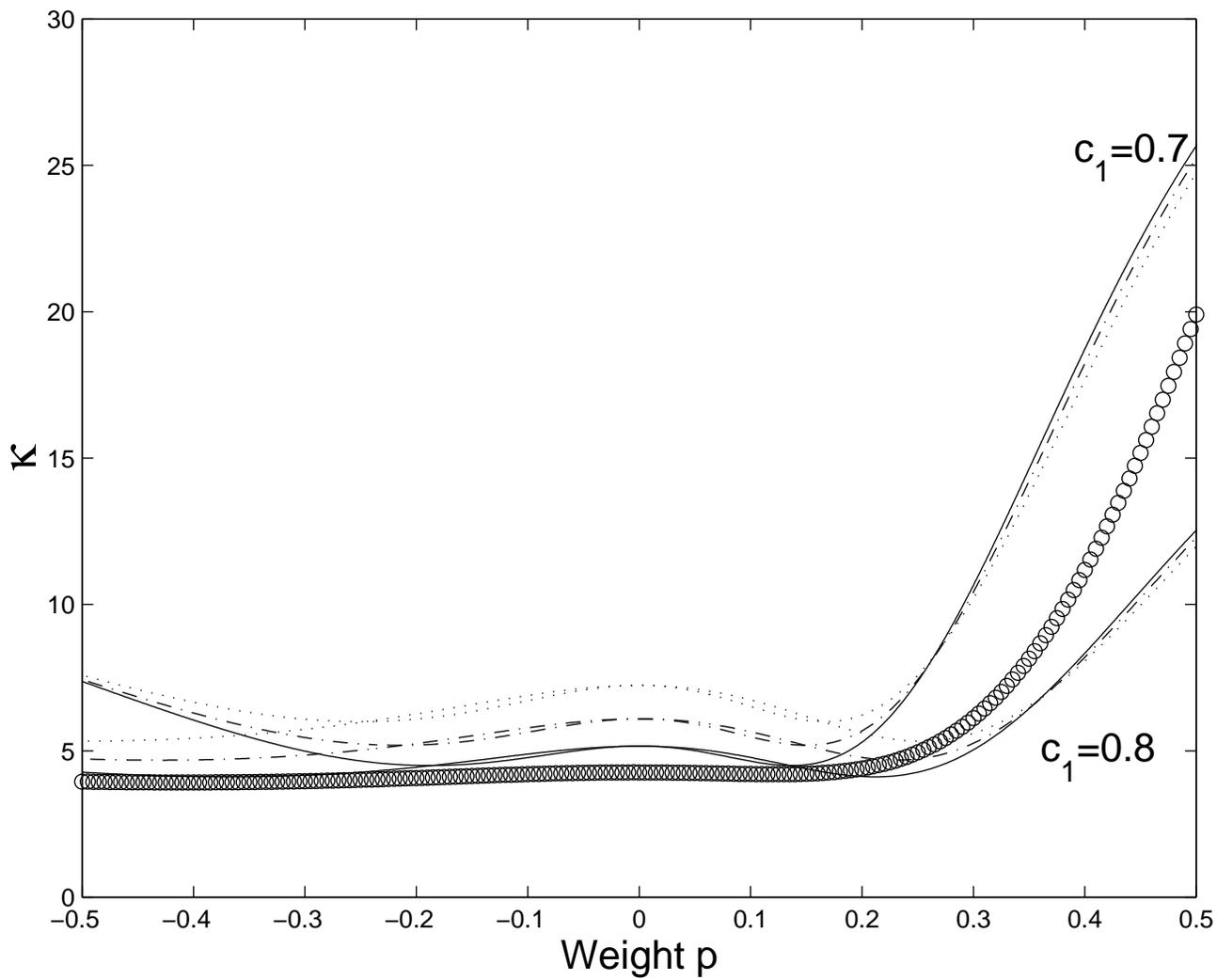
\end{document}